\documentclass[sigconf,nonacm]{acmart} 
\AtBeginDocument{%
  }

\setcopyright{none} 

\setcopyright{acmlicensed}
\copyrightyear{2026}
\acmYear{2026}
\acmDOI{XXXXXXX.XXXXXXX}
\acmConference[Conference acronym 'XX]{Make sure to enter the correct
  conference title from your rights confirmation email}{June 03--05,
  2018}{Woodstock, NY}
\acmISBN{978-1-4503-XXXX-X/2018/06}

\usepackage{tabularx}
\usepackage{array}
\newcolumntype{L}[1]{>{\raggedright\arraybackslash}p{#1}}

\begin{document}

\title{Rethinking How We Discuss the Guidance of Student Researchers in Computing}

\author{Shomir Wilson}
\email{shomir@psu.edu}
\orcid{https://orcid.org/0000-0003-1235-3754}
\affiliation{
  \institution{Pennsylvania State University}
  \city{University Park}
  \state{PA}
  \country{USA}
}


\begin{abstract}
Computing faculty at research universities are often expected to guide the work of undergraduate and graduate student researchers. This guidance is typically called \textit{advising} or \textit{mentoring}, but these terms belie the complexity of the relationship, which includes several related but distinct roles. I examine the \textit{guidance of student researchers in computing} (abbreviated to \textit{research guidance} or \textit{guidance} throughout) within a facet framework, creating an inventory of roles that faculty members can hold. By expanding and disambiguating the language of guidance, this approach reveals the full breadth of faculty responsibilities toward student researchers, and it facilitates discussing conflicts between those responsibilities. Additionally, the facet framework permits greater flexibility for students seeking guidance, allowing them a robust support network without implying inadequacy in an individual faculty member's skills. I further argue that an over-reliance on singular terms like \textit{advising} or \textit{mentoring} for the guidance of student researchers obscures the full scope of faculty responsibilities and interferes with improvement of those as skills. Finally, I provide suggestions for how the facet framework can be utilized by faculty and institutions, and how parts of it can be discussed with students for their benefit.
\end{abstract}




\maketitle

\section{Introduction}

Students often contribute to computing research at colleges and universities. Research is expected for most graduate degree programs, and undergraduates sometimes participate as well. The goal of a synergistic relationship is often implied or explicit in student involvement: the student learns from the research experience, especially in ways they cannot from coursework, and the student contributes to the creation of knowledge that benefits society. Guiding student research is also a way for computing faculty to bring together research and pedagogy for mutual benefit.

In contrast to the importance of guiding student researchers, faculty are typically given little or no training in this aspect of their work. In absence of training, they may rely on memories of how they were guided as student researchers, how they observe other faculty guiding student researchers, or piecemeal resources they find on their own. This lack of structure is consistent with a broader lack of training for university faculty---they are assumed to know how to teach and perform faculty-level research activities with a minimum of input---but it belies the complexity of the task.

Meanwhile, the latent heterogeneity of this guidance is underappreciated. Within it, faculty serve multiple roles.\footnote{Some possible faculty roles are listed here to establish the breadth of the problem. Their complex origins are documented in Section~\ref{mentorship_related}.} For example, as an \textit{advisor}, the faculty member is officially responsible for guiding a student's activities towards completing thesis requirements. As a \textit{mentor}, the faculty member guides a student's assimilation of professional norms. As a \textit{manager}, the faculty member hires the student---a formal agreement for labor---and gives them tasks, resources, and feedback. As a \textit{collaborator}, the faculty member works alongside the student on research. As a \textit{coach}, the faculty member helps a student gain and improve skills. As a \textit{sponsor}, the faculty member introduces the student to the professional community and helps to take advantage of opportunities. Overlap exists between these roles and others that the faculty member may have. However, neglecting the heterogeneity of this guidance promotes collapsing these roles into one, neglecting students' needs and conflicts of interest between the roles. This inattention is already part of the typical language used to discuss guidance: \textit{advising} is often spoken of as ambiguously including all those roles, some of those roles, or solely the official responsibilities recognized by the university.

This paper explores the overloaded meanings of \textit{advising} in the context of student research, using a facet framework to inventory several faculty roles that are related but different. \textit{Research guidance} or simply \textit{guidance} are proposed as terms to unite these roles, removing the ambiguity from \textit{advising} that obscures the breadth of faculty obligations. Interactions between roles are described, showing how the facet framework helps to understand conflicts of interest. Finally, the harms of an impoverished vocabulary for research guidance are discussed, leading to suggestions on how to discuss this guidance more productively.

\section{Related Work}
\label{related_work}


This section reviews prior work in two areas: guidance for student researchers in computing specifically and the broader realm of advising and mentorship of student researchers.

\subsection{Student Researchers in Computing}

Several prior works have focused on student researchers in computing. One area of emphasis has been engaging undergraduates in research. Sharma et al.~\cite{10.1145/3501385.3543976} studied faculty reluctance to work with undergraduate researchers, noting reasons that included undergraduates' limited time and misalignment of goals between the faculty member and the student. The sources of that misalignment become evident in the roles of the framework I present. Fernandez~\cite{10.1145/3641554.3701805, 10.1145/3626253.3635502} and Martin et al.~\cite{10.1145/3641555.3705118} explored using formal courses to train and motivate undergraduate researchers, showing the feasibility of direct overlap between faculty research guidance roles and teaching roles. Several works have shown the value of community-building among undergraduate researchers, which can promote student resilience~\cite{10.1145/3641555.3705214, 10.1145/3545945.3569841, 10.1145/3632620.3671104}; beyond the faculty roles that I examine in this paper, those prior works demonstrate the value of peer support. Others focused on the impact of undergraduate research on academic performance~\cite{10.1145/3626252.3630765, 10.1145/3545947.3576360}, and how the pandemic affected the learning assessment of undergraduate researchers~\cite{10.1145/3545947.3576327}. Beyond these prior works, Harrington et al.~\cite{10.1145/3641555.3705274} survey a large set of papers about undergraduate researchers in computing.

Graduate student researchers are less frequently studied, although some notable works center on their experiences. Broadening participation has received particular emphasis. Alm et al.~\cite{10.1145/3626253.3635489} focused on artificial intelligence research by graduate students, and Eicher et al.~\cite{10.1145/3641555.3705136} studied a framework to support online master's student researchers. Others examined the multi-nationalism of the graduate student research community, through distributed career mentoring~\cite{10.1145/3545945.3569789} and through soliciting graduate students' goals and unmet needs~\cite{10.1145/3545947.3576291}. Lunn et al.~\cite{10.1145/3545945.3569858} further used the international setting to study remote guidance of graduate student researchers~\cite{10.1145/3545945.3569858}. 

These works demonstrate the value of supporting student researchers, but the multiplicity of faculty roles remains outside their foci. Noteably, Alvarado et al.~\cite{10.1145/3408877.3432364} study how two mentors can provide complementary strengths, but these strengths are described mostly as mentoring. By directly examining the multiplicity of faculty roles, the framework I present also flexibly covers both undergraduate and graduate student researchers.

\subsection{Mentorship and Related Roles}
\label{mentorship_related}


\textit{Mentorship} and \textit{advising} are sometimes used as umbrella terms for all the guidance roles in this paper. Although I show how overloading these terms creates problems, to avoid misattribution, each prior work is described below using its preferred terminology. The extensive literature survey in Johnson and Griffin's \textit{On Being a Mentor}~\cite{johnson2024being} is noteworthy for partly informing this review.

Approaching guidance as a multifacted task has prior precedent. Levinson et al.~\cite{levinson1978seasons} describe mentorship in the context of adult development to include the roles of \textit{teacher} (in a way other sources may describe as \textit{coaching}~\cite{cavanagh2005evidence}) and \textit{sponsor}. Healy and Welchert~\cite{healy1990mentoring} and NASEM~\cite{nasem2019mentorship} describe mentorship with a professional-personal duality, including career advancement and ``identity transformation'' or ``psychosocial support'', respectively. Focusing on the educational context, ``faculty roles'' are frequently referenced but widely differ.  Santiesteban et al.~\cite{santiesteban2022defining} refer to \textit{advising}, \textit{coaching}, and \textit{mentoring} in the context of medical school students. AuCoin and Wright~\cite{aucoin2021student} surveyed students and found three themes in what they wanted from a faculty mentor: \textit{experience}, \textit{support/connection}, and \textit{professional development}. Others, such as Nesbit et al.~\cite{Nesbitt_Barry_Lawson_Diaz_2022}, acknowledge that faculty have multiple roles but do not enumerate them. Additionally, mentorship pyramids, where mentors and mentees operate in a hierarchy of seniority, have been studied~\cite{brannan2018mentoring, van2014towards}. While the helpfulness of this structure has been demonstrated, it does not illuminate the multiple roles of research guidance.

Finally, the guidance roles in this paper's framework are inspired by prior works, but those roles require re-synthesis to avoid vagueness and redundancies. \textit{Advisor}, \textit{manager}, \textit{coach}, and \textit{mentor} are described by Johnson and Griffin, but \textit{mentor} is sometimes used in their work as an umbrella term for the others and sometimes implied to be a peer to them. The framework resolves that ambiguity by considering all these roles as peers, and it adds two additional roles. \textit{Sponsor} appears informally in Johnson and Griffin's work~\cite{johnson2024being}, but the framework elevates it to a recognized role with inspiration from Jamison-McClung~\cite{jamison2022mentorship}. Describing support for women faculty, Jamison-McClung juxtaposed sponsorship and mentorship and described a sponsor as someone who ``nominate[s] and promote[s] colleagues for career-advancing opportunities and awards, both internally and externally''~\cite{jamison2022mentorship}. For \textit{collaborator}, I look to Katz and Martin's characterization~\cite{katz1997research}. They describe a broad set of collaboration scenarios that include working together on a project for its duration, co-writing the original research proposal, providing the overall project strategy, and contributing a key element to the work. Additionally, faculty-student coauthorship of manuscripts is normal within computing research~\cite{fernandes2022author}, making collaboration appropriate for inclusion.

\section{Context and Limitations}


Guidance norms vary between academic disciplines. Influences on those norms include student funding models (e.g., the relative availabilities of teaching assistantships, research assistantships, hourly wage positions, or other forms of support), research infrastructure (e.g., whether wet lab work or fieldwork are expected), and scholarly traditions (e.g., whether it is typical for faculty and their research advisees to coauthor papers). This section describes the context of computing research as relevant to the facet framework, distinguishing it from other disciplines. The applicability of observations in this paper may vary outside of computing, and even within it some variation is expected. Moreover, the norms described below are primarily from US institutions, while factors affecting guidance may vary widely in other countries. The authors acknowledge these limitations with the hope that the contents of this paper will still hold some relevance in other settings.

Student researchers in computing are frequently given hands-on research tasks to accomplish within teams. These include writing source code, collecting and curating datasets, running experiments, performing analysis on the results, and writing manuscripts. Importantly, there is \emph{collaboration} between the student and the faculty member: they are working together on research, and they are likely to coauthor. An alternative to this model, as potentially practiced in a different discipline, is that faculty members provide some guidance to student researchers but consider their research to be separate, leading to a different set of power dynamics in the relationship. When the faculty member supervises a student's research and coauthors with them, the faculty member has a greater investment in the success of that research, as part of building their professional portfolio and realizing their time and effort.

Related to the provision of hands-on tasks is the recognition that most student researchers will go on to careers outside academia. This is true for undergraduate researchers in any discipline, but for graduate researchers in certain disciplines (such as computing) it is distinctive, and it is especially pronounced for students in Ph.D. programs. In some disciplines academe is a likely career direction for Ph.D. graduates (e.g., much of the humanities~\cite{AmAcad2025HumanitiesPhDs}), but in computing it is common for Ph.D. graduates to find jobs in industry~\cite{CRA2025Taulbee}. This means many of the hands-on work habits that students develop as researchers could become relevant to their future jobs, making best practices in those habits especially relevant to their professional development.

Finally, faculty are expected to have a strong pedagogical influence on student researchers' activities. This is inherent in education as a setting, but juxtaposed with other features of the relationship, it becomes noteworthy: the faculty member must integrate guiding the student's education with supervising the student's work. These supervisory and pedagogical goals are not necessarily aligned. Circumstances may exist when guidance to maximize the student's research productivity is at odds with guidance to maximize what the student learns. Because research productivity benefits the faculty member differently than student learning, a conflict of interest arises. The difficulty of identifying those conflicts and the lack of recognized vocabulary for them motivates a principled examination of the faculty guidance of student researchers.

\section{Facets of Research Guidance}

\begin{figure}
    \centering \includegraphics[width=\columnwidth]{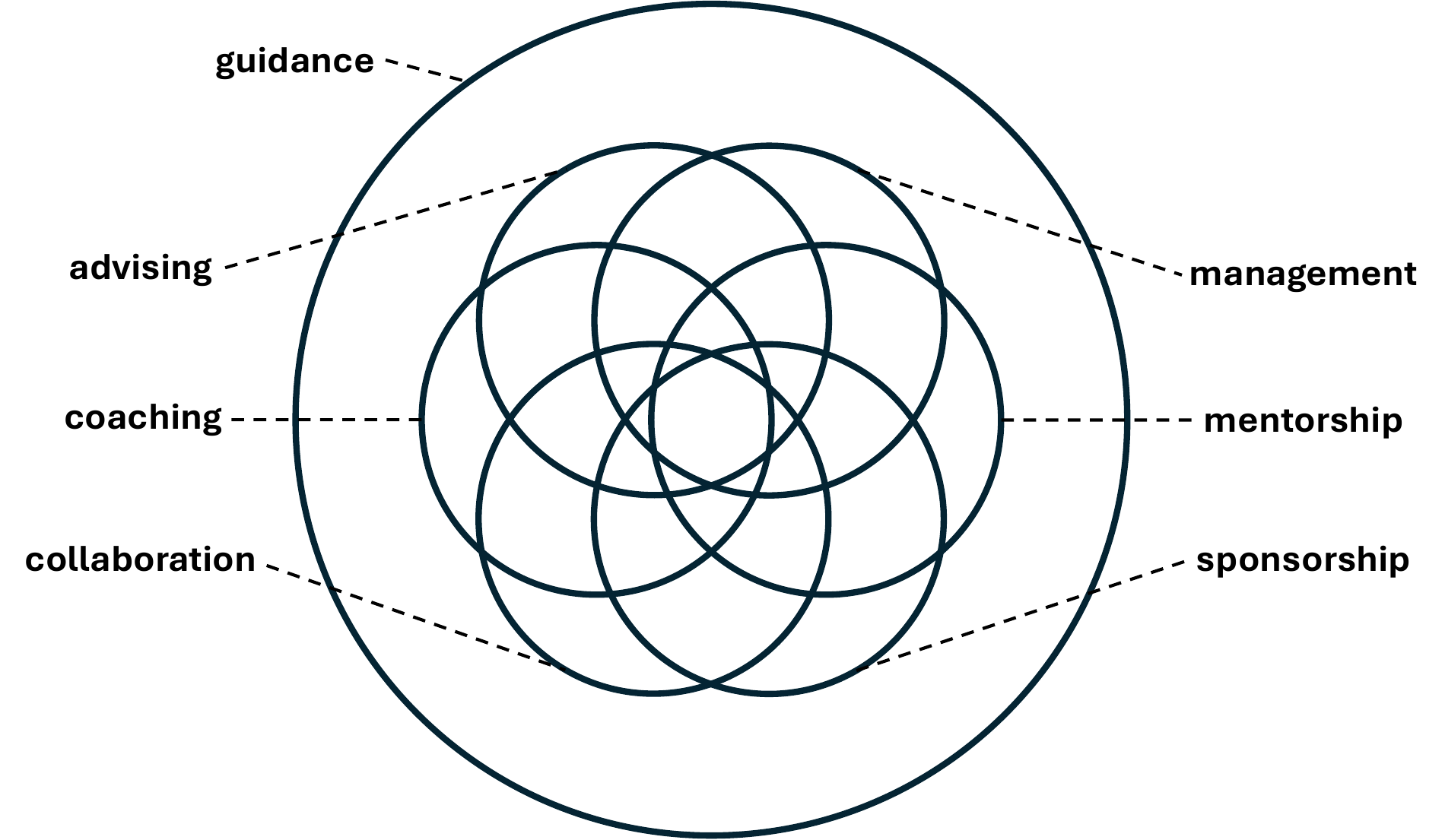}
    \caption{A Venn diagram showing \textit{guidance} as a broad concept that includes many overlapping but differing roles that faculty provide to student researchers.}
    \label{fig:facet}
\end{figure}


This section inventories some of the most common roles a faculty member can have in the guidance of a student researcher. These roles are presented in a \textit{facet framework}~\cite{guttman1971measurement, solomon2021introduction}. Within this framework, I show how terms such as \textit{advising} and \textit{mentorship} can be conceptualized more helpfully compared to their use in the status quo. Accordingly, I present \textit{guidance} as an all-encompassing term for the roles that a faculty member can have in a student researcher's education. Using this term reduces the ambiguity in how advising, mentorship, and related terms are discussed, allowing them to be contrasted within the framework.

Figure \ref{fig:facet} shows guidance to include six facets, which have territories that are partly independent of each other and partly overlapping. Because the figure is two-dimensional, only a subset of the possible overlap combinations are shown, but any combination is possible. At the center, the intersection of all facets represents support for the student's progress as a researcher. This set of facets is partly inspired by Johnson and Griffin's observations on advising, supervision\footnote{The framework uses \textit{manager} instead of \textit{supervisor} to lexically distance the concept from \textit{advisor}.}, coaching, and mentoring~\cite{johnson2024being}. However, Johnson and Griffin claim that mentorship is an outcome of the other three roles, not a role at parity with them, thus placing it in a different genus. While they are related, I argue below that independent mentorship is possible and even desirable. Johnson and Griffin's framework also uses \textit{mentorship} to describe the union of all roles, creating ambiguity. This framework adds to those four roles two more, \textit{sponsor} and \textit{collaborator}, with their origins and the motivations for adding them explained in Section~\ref{mentorship_related}.



\subsection{Inventory of Facets}

I describe each facet (or \textit{role}, interchangeably) in the framework. While other facets are possible, I use the prior work on these (as in Section~\ref{mentorship_related}) to focus upon them.


\subsubsection{Advisor} 

The faculty member is responsible for guiding a student's activities toward the completion of research-oriented degree or program requirements.\footnote{Another kind of advising exists, toward coursework requirements, but I set it aside because it is out of scope.} A typical requirement is a thesis or a scholarly paper, and I focus on that outcome for simplicity. However, in some settings (e.g., summer research programs) the requirement is a final presentation or report.

Advising is one of the most commonly discussed guidance roles, which matches its formal significance. The university recognizes the advisor-advisee relationship with policies and procedures, often including a thesis proposal and defense. These formal artifacts typically limit the advising role to one or two faculty per student, unlike some other guidance roles. In other words, a student typically can have only one or two advisors on record with their university, while they may have several mentors (a role described below). Additionally, advising is highly pedagogical and oriented solely around the student's benefit. 

Either a prospective advisor or prospective advisee may request this relationship, though (like all of the roles) mutual consent is necessary to proceed. The advisor's role is mainly to guide the student's work, a modality of instruction that differs subtly from the authority of a manager or the advice of a mentor or coach. The role ideally ends when the student completes their thesis, but either party can withdraw their consent earlier. The penalty for maladaption or separation prior to thesis completion is high: because the advisor's approval is the gateway to graduation, it may cost the student significant time and effort to continue without them.


\subsubsection{Mentor} 

The faculty member guides a student's assimilation of professional norms. These norms include the conceptual framework of research but also ethics, long-term goals, and other topics related to the conduct of a computing career. Research-active faculty are well-suited to mentor aspiring career researchers, but they also must guide mentees who seek careers outside of research.

Unlike the role of an advisor, this role has no formal product (i.e., a student's thesis) and it may not be formally recognized by the university. A student may have zero or more mentors, although having at least one is desirable. If the student has one mentor, their advisor is an intuitive candidate due to the nature of that role; however, it is also possible for the advisor-advisee relationship to be curtailed to the extent the advisor does not assume the role. The student may seek mentors in addition to (or in replacement of) their advisor. The number of mentors is limited chiefly by the available time and interest of the involved parties.

A mentor or mentee may initiate the relationship. Because of its informality, the initiation can be implicit (e.g., a request for help without mentioning \textit{mentorship}) and it can lack a well-defined beginning (e.g., a light conversation leading to progressively greater engagement over time). Like advising, the role is highly pedagogical and oriented around the student's development. The mentor may provide instructions, but they lack formal authority, which makes them more tenuous than instructions from an advisor or a manager. The role may end at any time when the mentor stops providing mentorship or the mentee stops seeking it, and that end may not be well-defined. Alternatively, the role may end in a \textit{de facto} sense when the student graduates---i.e., the faculty member can no longer mentor them as a student---leaving open the possibility of a transition to different forms of mentorship. The penalty for maladaption or separation prior to graduation is not formal like the loss of an advisor, and it does not prevent thesis completion. However, a lack of mentorship can pose developmental or emotional obstacles, as the student must navigate unspoken professional norms alone.


\subsubsection{Manager}

The faculty member hires the student into a research assistantship, gives the student tasks to perform, provides feedback on the student's work, and guides the student's participation in a workplace. The tasks of a research assistant are nominally research or research support, which permits significant overlap with the student's research toward their thesis. This double-counting is typically permitted, with the proviso that the overlap is incomplete: some research assistantship tasks are unresponsive to thesis requirements, and vice versa. A student's manager is often assumed to be their advisor, but exceptions are possible.

The manager-worker relationship is formally recognized by the university, which produces obligations to follow labor laws and to maintain a professional working environment. A student may have one or more managers, although when there are multiple managers the faculty members are obligated to coordinate and avoid conflicting instructions. Alternatively a student could have no manager, if they are not formally employed to perform research. This role is distinguished by its non-pedagogical basis, as a formal agreement for labor. It is possible for the goals of a manager to conflict with the goals of an advisor, a scenario I explore in Section~\ref{sec:conflicts}.

A student may request this relationship, but most of the burden to actualize it is on the faculty member. They must acquire financial resources to hire the student, evaluate the student's suitability for the position, and initiate the hiring procedure. Instructions from the manager are orders to perform tasks, and the student is expected to follow them to retain employment. The role can end for a variety of reasons: the student graduates, funds are no longer available to support the student's position, the student no longer needs (or wants) the position, or the faculty member is dissatisfied with the student's work performance. The penalty for maladaption is moderate to high: the student acquires basic workplace norms as a research assistant, and they may depend on the income from the position to continue toward their degree.


\subsubsection{Collaborator}

The faculty member works alongside the student on research. Both make intellectual contributions toward research outcomes, informally as research findings and formally as manuscripts, datasets, source code, or other artifacts. Unusually among the roles, the vocabulary for this one is symmetric: the faculty member is the collaborator of the student, and vice versa. An advisor or manager is likely to have this role in a student's research guidance, but other faculty members also may have it.

Whether the relationship is formal depends upon recognition in research artifacts. Artifacts that entail authorship make collaboration explicit, but the time to produce those artifacts varies widely, and until then, collaboration is an informal relationship. Among the guidance roles, this is one of two (along with \textit{coach}) that scale well: a student could have many faculty collaborators on a manuscript, although the degree of engagement with each collaborator can vary widely. For example, an advisor is likely to be more engaged than a collaborator who has no other relationship with a student and works at a different institution.

Collaboration requires cooperation, but other roles---particularly advisor and manager---can induce it to happen. The start of a collaboration can be formally acknowledged verbally or in writing, or the collaboration can develop in an \textit{ad hoc} manner over time. Collaboration does not entail one party giving orders to another, but power dynamics based upon relative seniority and other guidance roles can make instructions inevitable. The relationship ends when the research reaches a conclusion or either party withdraws. The penalty for maladaption can vary from mild (e.g., the student stops working with a non-advisor, non-manager collaborator who provided little input) to devastating (e.g., the student cannot continue with their thesis without input from any collaborators).


\subsubsection{Coach}

The faculty member helps a student gain and improve skills for their work. Student researchers must develop a variety of skills, such as presenting their research, writing well, performing a literature review, establishing good work habits, and acquiring technical skills within their discipline. An advisor, manager, or mentor can inherit the responsibilities of a coach from those respective roles, but other faculty also can have the role.

There are few formal opportunities to acknowledge this role, and it is highly situational: a faculty member can step into it when they see a need and exit quietly when the need has passed. Two examples of this transiency are a thesis committee member who provides feedback on the student's presentation or a faculty member whom the student contacts solely to discuss relevant methods for a project. The relationship is pedagogical, as the faculty member's contribution is advice and formative feedback. A student can have arbitrarily many sources of coaching, and some coaches may maintain that role for years while others have it for an hour or less. 

The start and end of a coaching relationship are \textit{ad hoc} like collaboration, but without the normative expectation that the relationship will last the duration of a project. The penalty for maladaption can vary from mild (e.g., the student stops seeking input from a specific faculty member and instead finds it elsewhere) to severe (e.g., a faculty member's coaching interferes with the student's progress).


\subsubsection{Sponsor}

The faculty member introduces the student to the professional community and to professional opportunities. The faculty member may connect the student to potential collaborators or employers, promote the student's achievements, forward fellowship solicitations and job openings to the student, or write reference letters. The advisor, mentor, or manager are likely to have this role, but other faculty members also can hold it. 

Reference letters are a formal acknowledgment of sponsorship, but some activities of sponsorship (such as the rest of those listed above) are informal and situational. An advisor may feel a natural obligation to sponsor a student for the duration of advising, but any other faculty member can help a student find and develop professional opportunities. For opportunities that require multiple references, such as applications for faculty positions, it is obligatory that the student seek multiple sponsors. The relationship is not pedagogical---other roles prepare a student to take advantage of opportunities---but it is solely intended to benefit the student.

Like collaboration and coaching, the start and end of sponsorship are \textit{ad hoc}. The importance of the relationship grows as the student begins engaging with their professional community and diminishes once the student establishes themselves in it. The penalty for maladaption can vary from mild (e.g., a student disengages from a problematic sponsor and finds help elsewhere) to severe (e.g., a faculty member damages the student's reputation).





\subsection{The Inadequacy of Any One Role}

\begin{table*}
\caption{A comparison of research guidance facets.}
\label{tab:comparison}
\begin{tabular}{>{\bfseries}L{0.12\textwidth} | L{0.12\textwidth} | L{0.12\textwidth} | 
                L{0.12\textwidth} | L{0.12\textwidth} | L{0.12\textwidth} | 
                L{0.12\textwidth}}
 & \textbf{Advisor} & \textbf{Mentor} & \textbf{Manager} & \textbf{Collaborator} & \textbf{Coach} & \textbf{Sponsor} \\
\hline
Formal Recognition & By the academic program & Possibly none & By the university, as employment & By co-authorship & Possibly none & Documentation as a reference, or possibly none \\
\hline
Pedagogical Value & Entirely pedagogical & Entirely pedagogical & Indirect & Indirect & Entirely pedagogical & Not pedagogical \\
\hline
Quantity Per Student & Typically 1-2 & Limited by time or interest & Typically 1-2 & Limited by project size & Highly situational, unlimited & Limited by availability or need \\
\hline
Starting Point & Faculty member agrees to advise student & Possibly explicit or implicit & Faculty member hires student & Commencement of work & Loosely defined & Loosely defined \\
\hline
Finish Line & Student graduates & Student graduates, or beyond & Student graduates or position expires & Project completion & Student acquires relevant skills & Student's career is launched \\
\end{tabular}
\end{table*}

Table \ref{tab:comparison} reviews the roles. A side-by-side comparison shows how any one role is insufficient to represent all the others. Formal recognition varies from obligatory (\textit{advisor} and \textit{manager}) to \textit{ad hoc} (\textit{collaborator}) to possibly nonexistent (\textit{mentor}, \textit{coach}, \textit{sponsor}). Pedagogical value is typically assumed when faculty interact with students, and the context of higher education makes pedagogy paramount, but it is indirect for some roles (\textit{manager} and \textit{collaborator}) and a misnomer for \textit{sponsor}. Quantity per student is limited to typically one or two for formally recognized roles, but the number is open-ended for others. Starting points vary from rigidly to loosely defined, and some goals can have finish lines long before graduation without detriment to the student. (Note the term \textit{finish line} here exempts situations when the relationship ends before meeting its goals.) 

\textit{Advisor} and \textit{mentor} are commonly used as umbrella terms for guidance roles, as I showed by reviewing prior literature in Section~\ref{related_work}. However, neither is suitable to replace all of its peers in the facet framework. Immediately between \textit{advisor} and \textit{mentor}, a mismatch occurs: \textit{advisor} receives institutional recognition and \textit{mentor} does not require that recognition. A faculty member can mentor a student without being their advisor, and it is considered normal or beneficial for a mentee to have multiple mentors~\cite{montgomery2018mentoring}. Conversely, it is possible for a faculty member to serve as an advisor without being a mentor, if contact between them is (perhaps problematically) limited to the satisfaction of degree requirements. The \textit{manager} role has wide formal recognition (often as \textit{supervisor}), but it is also inadequate to cover all guidance roles. The framework of a labor agreement alone does not create pedagogical benefit, and the role relies on the others to induce that benefit.

These differences between roles suggest that ambiguity among them leads to problems. In discussions of guidance, if two participants have different notions of scope for these terms, they may not realize the lack of agreement. For example, when an academic administrator asks a faculty member to increase their mentoring, the two may interpret that request differently. Alternatively, a faculty member who is unaware of some guidance roles---e.g., if they solely focus on advising and managing---may not understand a student's obstacles to success. Recognizing these roles also facilitates examining conflicts between them, which I discuss next.

\subsection{Conflicts Between Roles}
\label{sec:conflicts}

I explore some conflicts between roles as examples, while acknowledging that others likely exist. In scenarios below, unless otherwise specified, a faculty member can hold an arbitrary number of roles (and possibly all of them). The roles also can be distributed among multiple faculty members.

Tension exist between roles that focus on productivity and those that focus on pedagogy. \textit{Manager} and \textit{collaborator} rely on the student's research output for mutual gain, in the sense that both the student and the faculty member benefit from the student's work. However, \textit{advisor} and \textit{mentor} are (ideally) closer to selfless, as they are aligned with the student's degree completion and professional development, respectively. A selfless advisor and mentor are obligated to encourage a productive student to complete their degree requirements and graduate promptly, while a manager or collaborator benefits by the student staying longer. 

Even when considering productivity or pedagogy alone, tensions still arise. For productivity, when \textit{manager} and \textit{collaborator} are held respectively by different faculty, there exist scenarios when moving the student to a new project benefits the manager while retaining the student on an ongoing project benefits the collaborator. For pedagogy, there exist scenarios when an \textit{advisor} sees a way for a student to make rapid degree progress while a \textit{mentor} perceives a need to let the student choose among options. In these cases and others, single-role frameworks provide a less nuanced picture of the conflict, or they do not reveal the conflict at all.

\section{Praxis}

I provide suggestions for how faculty and institutions can use this facet framework, and I also touch upon how students can be exposed to the framework for their benefit.

\label{sec:for_faculty}
\subsection{For Faculty and Institutions}

Some of the terms in the facet framework are already widely used, even if ambiguously, and overriding status quo beliefs about them (e.g., the belief that \textit{mentorship} is chiefly accomplished by \textit{advising}) is a difficult task. However, awareness of the multiplicity of roles is a step toward more nuanced, fuller guidance of student researchers. When faculty are given advice or feedback on guidance, describing roles like those in the framework---or even a carefully chosen subset of them---can encourage faculty to think about their work with student researchers as multifaceted and worthy of further reflection. As I show, the term \textit{guidance} appears to be suitable to encompass all roles without overloading existing terms.

The framework also can be used to encourage faculty to acknowledge the value of a robust guidance network for their students, including support for the student from colleagues and collaborators. The status quo connotations of \textit{advising} and \textit{mentorship} (here as an imprecise synonym for \textit{advising}) are monolithic, promoting the view that a faculty member can or must satisfy all the student's needs for guidance. This view can subtly limit the guidance a faculty member gives to students. Instead, knowing the diversity of roles that faculty perform for student researchers, as well as how some roles scale to involvement by several faculty, makes it apparent why engaging with multiple faculty is appropriate for a student researcher and even obligatory. In this distributed model, the \textit{advisor} or \textit{manager} can think of themselves as the \textit{anchor} for the student's guidance, by virtue of the official origins of those roles. 

Finally, by separating productivity- and pedagogy-focused roles, the framework allows faculty to reason through decisions when they feel conflicted about guiding a student. Pedagogy is implicit to the context of student research, which means that productivity sometimes must be sacrificed for the student's benefit. While the primacy of benefiting students has an immense intuitive basis, reasoning through it with a framework may appeal to faculty who feel more comfortable having detailed reasons for making tradeoffs in research productivity. The framework also provides the vocabulary for discussing such tradeoffs with colleagues or administrators.

\subsection{For Guiding Student Researchers}



Sharing the framework with students can help them recognize the kinds of help they can ask for and give them language to express their needs. Additionally, situational sharing of relevant parts of the framework is likely to benefit student researchers. This sharing can take multiple forms. One of them is transparency for difficult decisions: when a faculty member explains how they reason through conflicting guidance roles, the student gains an appreciation for how to navigate similar problems. (This sharing is itself a form of mentoring.) This appreciation is immediately useful for graduate students who mentor undergraduates, and it becomes relevant later to students who are future faculty members.

Another appropriate time to expose parts of the framework to students is when recommending to a student that they seek additional guidance from other faculty. Naming and explaining the relevant role can help the student frame their requests to other faculty members (e.g., that they seek coaching on a specific skill or mentorship toward a professional goal). Unambiguous requests save time and effort for both students and faculty, promoting successful interaction. Additionally, faculty can make themselves approachable by explaining in publicly-available advising statements how they fulfill guidance roles, regardless of whether those roles are explicitly named. This openness is consistent with the idea that a ``hidden curriculum'' exists in postsecondary education~\cite{wilson2024documenting}, and students perform better when they are exposed to it.

\section{Conclusion}

I presented a facet framework for the roles faculty possess when guiding student researchers in computing. While more facets are possible, the six described in this manuscript receive frequent discussion and have a basis in prior literature. Importantly, this itemization of roles removes ambiguity in the status quo terminology, which potentially obscures faculty responsibilities during discussions of student research. The framework helps to identify and resolve conflicts between faculty roles, and it provides a basis for directing faculty toward robust guidance of student researchers.

\begin{acks}

This manuscript is based upon work supported by the National Science Foundation under Grant No. 2237574. I also thank my student researchers for their work, inspiring me to write this manuscript.

\end{acks}

\bibliographystyle{ACM-Reference-Format}
\bibliography{sigcse_2026_shomir}

\end{document}